\documentclass[final,4p,authoryear]{elsarticle}

\usepackage{amssymb}
\usepackage{amsmath}
\usepackage{bbm}
\usepackage{booktabs}
\usepackage{caption}

\newcommand\bbone{\ensuremath{\mathbbm{1}}}
\newcommand{\suchthat}{\;\ifnum\currentgrouptype=16 \middle\fi|\;}
\def\qmo{``}
\def\qmc{''}
\def\qmcsp{'' }

\def\bbeta{\mbox{\boldmath $\beta$}}

\def\bphi{\mbox{\boldmath $\phi$}}

\def\btheta{\mathbf{\theta}}

\def\0{\mbox{\bf{0}}}

\def\bx{\mathbf{x}}

\def\sT{\mathsf{T}}

\def\sN{\mathsf{N}}

\newcommand\url{\texttt}

\begin{document}

\begin{frontmatter}



\title{\LARGE Comparison of Value--at--Risk models:\\ the MCS package}


\author[uniroma1-memotef]{M. Bernardi}
\author[uniroma2-def]{L. Catania}
\address[uniroma1]{University of Padua, Department of Statistical Sciences}
\address[uniroma2-def]{University of Rome, Tor Vergata, Department of Economics and Finance}

\begin{abstract}
This paper compares the Value--at--Risk (VaR) forecasts delivered by alternative model specifications using the Model Confidence Set (MCS) procedure recently developed by \cite{hansen_etal.2011}. The direct VaR estimate provided by the Conditional Autoregressive Value--at--Risk (CAViaR) models of \cite{engle_manganelli.2004} are compared to those obtained by the popular Autoregressive Conditional Heteroskedasticity (ARCH) models of \cite{engle.1982} and to the recently introduced Generalised Autoregressive Score (GAS) models of \cite{creal_etal.2013} and \cite{harvey.2013}.
The Hansen's procedure consists on a sequence of tests which permits to construct a set of \qmo superior\qmcsp models, where the null hypothesis of Equal Predictive Ability (EPA) is not rejected at a certain confidence level. Our empirical results, suggest that, after the Global Financial Crisis (GFC) of 2007--2008, highly non--linear volatility models deliver better VaR forecasts for the European countries as opposed to other regions. The \textsf{R} package \texttt{MCS} is introduced for performing the model comparisons whose main features are discussed throughout the paper.
\end{abstract}


\end{frontmatter}
%
\section{Introduction}
\label{sec:intro}
%
\noindent During last decades hundred of models have been developed, estimated and validated from both an empirical and theoretical perspective. As a result, several alternative model specifications are usually available to the econometricians to address the same empirical problem. Just to confine our considerations within a given family, and without claiming to be complete, the Autoregressive Conditional Heteroskedastic (ARCH) models of \cite{engle.1982} and \cite{bollerslev.1986}, for example, have seen an exponentially increasing number of different specifications in the last few decades. Despite their popularity, they do not exhaust the set of models introduced for dynamic conditional volatility modelling which includes also the stochastic volatility models initially proposed by \cite{taylor.1994} and extensively studied by \cite{harvey_shephard.1996} and \cite{gallant_etal.1997} within the context of non--linear state space models. The family of dynamic conditional volatility models has been recently enlarged by the GAS model of \cite{harvey.2013} and \cite{creal_etal.2013} also known as Dynamic Conditional Score (DCS). The availability of such an enormous number of models raises the question of providing a statistical method or procedure that delivers the \qmo best\qmcsp model with respect to a given criterium. Furthermore, a model selection issue appears when the usual comparing procedures does not deliver an unique result. This could happen for example, when models are compared in terms of their predictive ability, so that models that produce better forecasts are preferred. Unfortunately, when evaluating the performances of different forecasting models it is not always trivial to establish which one clearly outperforms the remaining available alternatives. This problem is particularly relevant even from an empirical perspective especially when the set of competing alternatives is large. As observed by \cite{hansen_lunde.2005} and \cite{hansen_etal.2011}, it is unrealistic to expect that a single model dominates all the competitors either because the different specifications are statistically equivalent or because there is not enough information coming from the data to univocally discriminate the models.\newline
\indent Recently, several alternative testing procedures have been developed to deliver the \qmo best fitting\qmcsp model, see e.g. the Reality Check (RC) of \citet{white.2000}, the Stepwise Multiple Testing procedure of \cite{romano_wolf.2005}, the Superior Predictive Ability (SPA) test of \citet{hansen_lunde.2005} and the Conditional Predictive Ability (CPA) test of \cite{giacomini_white.2006}.
Among those multiple--testing procedures, the Model Confidence Set procedure (MCS) of \cite{hansen_etal.2003}, \cite{hansen_reinhard.2005} and \cite{hansen_etal.2011} consists of a sequence of statistic tests which permits to construct, the \qmo Superior Set of Models\qmcsp (SSM), where the null hypothesis of equal predictive ability (EPA) is not rejected at certain confidence level $\alpha$. The EPA statistic test is evaluated for an arbitrary loss function, which essentially means that it is possible to test models on various aspects depending on the chosen loss function. The possibility to specify user supplied loss functions enhances the flexibility of the procedure that can be used to test several different aspects. The MCS procedure starts from an initial set of $m$ competing models, denoted by $\mathrm{M}^{0}$, and results in a (hopefully) smaller set of superior models, the SSM, denoted by $\mathrm{\widehat{M}}_{1-\alpha}^{*}$. Of course, the best scenario is when the final set consists of a single model. At each iteration, the MCS procedure tests the null hypothesis of EPA of the competing models and ends with the creation of the SSM only if the null hypothesis is accepted, otherwise the MCS is iterated again and the EPA is tested on a smaller set of models obtained by eliminating the worst one at the previous step.\newline
\indent This paper compares the Value--at--Risk (VaR) forecasts delivered by alternative model specifications recently introduced in the financial econometric literature, using the MCS procedure, similarly to \cite{caporin_mcaleer.2014} and \cite{chen_gerlach.2013}. More specifically, the direct quantile estimates obtained by the dynamic CAViaR models of \cite{engle_manganelli.2004} are compared with the VaR forecasts delivered by several ARCH--type models of \cite{engle.1982} and \cite{bollerslev.1986} and with those obtained by two different specifications of the GAS models of \cite{creal_etal.2013} and \cite{harvey.2013}. The CAViaR model of \cite{engle_manganelli.2004} has been proven, see e.g. \cite{chen_gerlach.2012}, to provide reliable quantile--based VaR estimates in several empirical experiments. During the last few decades, the ARCH--type models of \cite{bollerslev.1986} have became a standard approach to model the conditional volatility dynamics. Those approaches are compared to the new class of score driven models, which are promising in modelling highly nonlinear volatility dynamics, see e.g., \cite{creal_etal.2013} and \cite{harvey.2013}. Up to our knowledge, VaR forecasting comparisons using the MCS procedure has not been considered using nonlinear GAS dynamic models.
\noindent Our empirical results, suggest that, after the Global Financial Crisis (GFC) of 2007--2008, highly non--linear volatility models deliver better VaR forecasts for the European countries. On the contrary, for the North America and Asia Pacific regions we find quite homogeneous results with respect to the models' complexity. As discussed in \cite{kowalski_shachmurove.2014}, this empirical finding is consistent with the greater impact that the GFC had in the European financial markets as compared to the non--Euro areas.\newline
\indent The models belonging to the superior set delivered by the Hansen's procedure can then be used for different purposes. For example, they can be used to forecast future volatility levels, to predict the future observations, conditional to the past information, or to deliver future Value--at--Risk estimates, as argued by \cite{bernardi_etal.2014}. Alternatively, the models can be combined together to obtain better forecast measures. Since the original work of \cite{bates_granger.1969}, a lot of papers have argued that combining predictions from alternative models often improves upon forecasts based on a single \qmo best\qmcsp model. In an environment where observations are subject to structural breaks and models are subject to different levels of misspecification, a strategy that pools information coming from different models typically performs better than methods that try to select the best forecasting model. Our analysis also considers the Dynamic Model Averaging technique proposed by \cite{bernardi_etal.2014} in order to aggregate the VaR forecasts delivered by the SSM, conditional on model's past out--of--sample performances as in \cite{samuels_sekkel.2011} and \cite{samuels_etal.2013}. For further information about the application of the model averaging methodology the reader is referred to \cite{bernardi_etal.2014}. Our results confirm that, under an optimal combination of models, individual VaR forecasts can be substantially improved with respect to standard backtest measures.\newline
\indent Another relevant contribution of the paper concerns the development of the \textsf{R} \citep{R.2013} package \texttt{MCS}. The \texttt{MCS} package provides an integrated environment for the comparison of alternative models or model's specifications within the same family using the MCS procedure of \cite{hansen_etal.2011} and it is available on the \texttt{CRAN} repository, \url{http://cran.r-project.org/web/packages/MCS/index.html}.
The \texttt{MCS} package is very flexible since it allows for the specification of the model's types and loss functions that can be supplied by the user. This freedom allows for the user to concentrate on substantive issues, such as the construction of the initial set of model's specifications, $\mathrm{M}^{0}$, without being limited by the constraints imposed by the software.\newline
\indent The layout of the paper is as follows. In Section \ref{sec:mcs_procedure} we present the \cite{hansen_etal.2011}'s MCS procedure highlighting the alternative specifications of the test statistics. Section \ref{sec:model_specifications} details about the alternative model specifications used to compare VaR forecasts. Section \ref{sec:mcs_package} presents the main features of the \texttt{MCS} package. Section \ref{sec:application} covers the empirical application which also aims at illustrating the implementation of the procedure using the provided package. Section \ref{sec:conclusion} concludes the paper.
%
\section{The MCS procedure}
\label{sec:mcs_procedure}
%
\noindent The availability of several alternative model specifications being able to adequately describe the unobserved data generating process (DGP) opens the question of selecting the \qmo best fitting model\qmcsp according to a given optimality criterion. This paper implements the MCS procedure recently developed by \cite{hansen_etal.2011} with the aim to compare the VaR forecasts obtained by several alternative model specifications. The Hansen's procedure consists of a sequence of statistic tests which permits to construct a set of \qmo superior\qmcsp models, the \qmo Superior Set Models\qmcsp (SSM), where the null hypothesis of equal predictive ability (EPA) is not rejected at a certain confidence level $\alpha$. The EPA statistic tests is calculated for an arbitrary loss function that satisfies general weak stationarity conditions, meaning that we could test models on various aspects, as, for example, punctual forecasts, as in \cite{hansen_lunde.2005}, or in--sample goodness of fit, as in \cite{hansen_etal.2011}. Formally, let $y_t$ denote the observation at time $t$ and let $\hat{y}_{i,t}$ be the output of model $i$ at time $t$, the loss function $\ell_{i,t}$ associated to the $i$--th model is defined as
\begin{equation}
\ell_{i,t}=\ell\left(y_t,\hat{y}_{i,t}\right),
\label{eq:loss}
\end{equation}
and measures the difference between the output $\hat{y}_{i,t}$ and the \qmo a posteriori\qmcsp realisation $y_t$. As an example of loss function, \cite{bernardi_etal.2014} consider the asymmetric VaR loss function of \cite{gonzalez_etal.2004} to compare the ability of different GARCH specifications to predict extreme loss in high frequency return data. The asymmetric VaR loss function of \cite{gonzalez_etal.2004} is defined as
\begin{equation}
\label{eq:loss_function_asymmetric}
\ell\left(y_{t},{\rm VaR}_{t}^{\tau}\right)=\left(\tau-d_{t}^{\tau}\right)\left(y_{t}-\mathrm{VaR}_{t}^{\tau}\right),
\end{equation}
where $\mathrm{VaR}_{t}^{\tau}$ denotes the $\tau$--level predicted VaR at time $t$, given information up to time $t-1$, $\mathcal{F}_{t-1}$, and $d_{t}^{\tau}=\bbone\left(y_{t}<\mathrm{VaR}_{t}^{\tau}\right)$ is the $\tau$--level quantile loss function. The asymmetric VaR loss function represents the natural candidate to backtest quantile--based risk measures since it penalises more heavily observations below the $\tau$--th quantile level, i.e. $y_t<\mathrm{VaR}^{\tau}_t$. Details about the loss function specifications can be found in \cite{hansen_lunde.2005} and in the following Section \ref{sec:loss_functions}.\newline
\indent We now briefly describe how the MCS procedure is implemented. The procedure starts from an initial set of models $\mathrm{M}^{0}$ of dimension $m$, encompassing all the alternative model specifications, and delivers, for a given confidence level $\alpha$, a smaller set, the superior set of models (SSM), $\mathrm{\hat{M}}_{1-\alpha}^{*}$, of dimension $m^*\leq m$. The SSM, $\mathrm{\hat{M}}_{1-\alpha}^{*}$, contains all the models having superior predictive ability according to the selected loss function. Of course, the best scenario is when the final set consists of a single model, i.e. $m^*=1$. Formally, let $d_{ij,t}$ denote the loss differential between models $i$ and $j$:
\begin{equation}
d_{ij,t}=\ell_{i,t}-\ell_{j,t},\quad i,j=1,\dots,m,\quad t=1,\dots,n,
\end{equation}
and let
\begin{equation}
d_{i\cdot,t}=\left(m-1\right)^{-1}\sum_{j\in\mathrm{M}\setminus\left\{i\right\}}d_{ij,t},\quad i=1,\dots,m,
\end{equation}
be the average loss of model $i$ relative to any other model $j$ at time $t$. The EPA hypothesis for a given set of models ${\rm M}$ can be formulated in two alternative ways:
\begin{align}
\mathrm{H}_{0,{\rm M}} & :  c_{ij}=0,\qquad\mathrm{for~all}\quad i,j=1,2,\dots,m\nonumber\\
\mathrm{H}_{{\rm A},{\rm M}} & :  c_{ij}\neq 0,\qquad\mathrm{for~some}\quad i,j=1,\dots,m,
\label{eq:EPA1}
\end{align}
or
\begin{align}
\mathrm{H}_{0,{\rm M}} & :  c_{i\cdot}=0,\qquad\mathrm{for~all}\quad i=1,2,\dots,m\nonumber\\
\mathrm{H}_{{\rm A},{\rm M}} & :  c_{i\cdot}\neq 0,\qquad\mathrm{for~some}\quad i=1,\dots,m,
\label{eq:EPA2}
\end{align}
where $c_{ij}=\mathbb{E}\left(d_{ij}\right)$ and $c_{i\cdot}=\mathbb{E}\left(d_{i\cdot}\right)$ are assumed to be finite and time independent.
According to \citet{hansen_etal.2011}, the two hypothesis defined in equations \eqref{eq:EPA1}--\eqref{eq:EPA2} can be tested by constructing the following two statistics
\begin{align}
\label{eq:epa_statistic_1}
t_{ij}&=\frac{\bar d_{ij}}{\sqrt{\widehat{\rm var}\left(\bar d_{ij}\right)}},\\
\label{eq:epa_statistic_2}
t_{i\cdot}&=\frac{\bar d_{i,\cdot}}{\sqrt{\widehat{\rm var}\left(\bar d_{i,\cdot}\right)}},
%
\end{align}
for $i,j\in\mathrm{M}$, where $\bar{d}_{i,\cdot}=\left(m-1\right)^{-1}\sum_{j\in\mathrm{M}\setminus\left\{i\right\}}\bar d_{ij}$ is the average loss of the $i$--th model relative to the average losses across the models belonging to the set $\mathrm{M}$, and $\bar{d}_{ij}=m^{-1}\sum_{t=1}^m d_{ij,t}$ measures the relative average loss between models $i$ and $j$. The variances $\widehat{\rm var}\left(\bar d_{i,\cdot}\right)$ and $\widehat{\rm var}\left(\bar d_{ij}\right)$ are bootstrapped estimates of ${\rm var}\left(\bar d_{i,\cdot}\right)$ and ${\rm var}\left(\bar d_{ij}\right)$, respectively. The bootstrapped variances $\widehat{\rm var}\left(\bar d_{i,\cdot}\right)$ and $\widehat{\rm var}\left(\bar d_{ij}\right)$ are calculated by performing a block--bootstrap procedure where the block length $p$ is set as the maximum number of significants parameters obtained by fitting an AR$\left(p\right)$ process on the $d_{ij}$ terms. In the \texttt{MCS} package the user can either specify an arbitrary block length $p$ or use the default AR$\left(p\right)$ procedure.
%
%
Details about the implemented bootstrap procedure can be found in \cite{white.2000}, \cite{kilian.1999}, \cite{clark_mccracken.2001}, \cite{hansen_etal.2003}, \cite{hansen_lunde.2005}, \cite{hansen_etal.2011} and \cite{bernardi_etal.2014}.
The statistic $t_{ij}$ is used in the well know test for comparing two forecasts; see e.g., \citet{diebold_mariano.2002} and \citet{west.1996}, while the second one is used in \citet{hansen_etal.2003}, \cite{hansen_reinhard.2005}, \cite{hansen_etal.2011}.
As discussed in \cite{hansen_etal.2011}, the two EPA null hypothesis presented in equations \eqref{eq:EPA1}--\eqref{eq:EPA2} map naturally into the two test statistics
\begin{equation}
\mathrm{T}_{\rm R,M}=\max_{i,j \in\mathrm{M}}\mid t_{ij}\mid\quad\mathrm{and}\quad\mathrm{T}_{\max ,\rm M}=\max_{i \in \mathrm{M}}t_{i\cdot},
\label{eq:epa_test_stat}
\end{equation}
where $t_{ij}$ and $t_{i.}$ are defined in equation \eqref{eq:epa_statistic_1}--\eqref{eq:epa_statistic_2}. Since the asymptotic distributions of the two test statistics is nonstandard, the relevant distributions under the null hypothesis is estimated using a bootstrap procedure similar to that used to estimate ${\rm var}\left(\bar d_{i,\cdot}\right)$ and ${\rm var}\left(\bar d_{ij}\right)$. For further details about the bootstrap procedure, see e.g., \citet{white.2000}, \cite{hansen_etal.2003}, \cite{hansen_reinhard.2005}, \cite{kilian.1999} and \cite{clark_mccracken.2001}.\newline
\indent As said in the Introduction, the MCS procedure consists on a sequential testing procedure, which eliminates at each step the worst model, until the hypothesis of equal predictive ability (EPA) is accepted for all the models belonging to the SSM. At each step, the choice of the worst model to be eliminated has been made using an elimination rule that is coherent with the statistic test defined in equations \eqref{eq:epa_statistic_1}--\eqref{eq:epa_statistic_2} which are
\begin{equation}
e_{\rm R,M}=\arg\max_{i}\left\{\sup_{j \in \rm M}\frac{\bar{d}_{ij}}{\sqrt{\widehat{\rm var}\left(\bar d_{ij}\right)}}\right\},\qquad
e_{\max,\rm M}=\arg\max_{i\in\rm M}\frac{\bar{d}_{i,\cdot}}{\sqrt{\widehat{\rm var}}\left(\bar d_{i,\cdot}\right)}
,
\label{eq:hansen_test_elimination_rule}
\end{equation}
respectively.
Summarazing, the MCS procedure to obtain the SSM, consists of the following steps:
\begin{itemize}
\item[1.] set $\mathrm{M}=\mathrm{M}^{0}$;
\item[2.] test for EPA--hypothesis: if EPA is accepted terminate the algorithm and set $\mathrm{M}_{1-\alpha}^{*}=\mathrm{M}$, otherwise use the elimination rules defined in equation \eqref{eq:hansen_test_elimination_rule} to determine the worst model;
\item[3.] remove the worst model, and go to step 2.
\end{itemize}
%
%
Since the Hansen's procedure usually delivers a SSM, $\hat{\rm M}_{1-\alpha}^*$, which contains a large number of models, in the next sections, we also describe how to implement a procedure that combines the VaRs forecasts.
%
\section{Model specifications}
\label{sec:model_specifications}
%
\noindent In our empirical illustration we apply the MCS procedure detailed in Section \ref{sec:mcs_procedure} to compare the VaR forecasts obtained by fitting a list of popular models introduced in the econometric literature over the last few decades. The autoregressive conditional heteroskedastic models, introduced by \cite{engle.1982} and \cite{bollerslev.1986}, are probably among the most widely employed tools in quantile--based quantitative risk management. Here, we consider the ARCH--type models not only because of their popularity, but also because of their ability to account for the main stylised facts about financial returns. Moreover, since the seminal paper of \cite{engle.1982}, hundreds of different specifications have been proposed, some of them having a huge flexibility in handling series having different characteristics. Despite their vast popularity, the ARCH models are principally focused on the scale of the modelled conditional distributions. The GAS models, introduced by \cite{creal_etal.2013} and \cite{harvey.2013}, have been recently gaining popularity also because they nest some of the traditional dynamic conditional variance approaches enlarging the class of time--varying parameter models. For the purposes of this paper, the ARCH family of models is discussed in Section \ref{sec:arch_models}, while GAS models are considered in Section \ref{sec:gas_models}. ARCH--type and GAS models provide an indirect estimation of the quantile--based risk measures because of the parametric assumption of the conditional distribution. A way to overcome the need to specify the conditional distribution is to model directly the quantile as in the Conditional Autoregressive Value--at--Risk (CAViaR) approach of \cite{engle_manganelli.2004}. The CAViaR specifications are briefly discussed in Section \ref{sec:caviar_models}.\newline
\indent Let $y_t$ be the logarithmic return at time $t$, throughout the paper, we consider the following general formulation encompassing all the considered model specifications
\begin{align}
\label{eq:general_model_spec_meas}
y_t\mid\left(\mathcal{F}_{t-1},\zeta_t,\vartheta\right)&\sim\mathcal{D}\left(y_t,\zeta_t,\vartheta\right),\qquad t=1,2,\dots,T,\\
\label{eq:general_model_spec_trans}
\zeta_t&=h\left(\zeta_{t-1},\dots,\zeta_{t-p},y_{t-1},\dots,y_{t-p},\mid\mathcal{F}_{t-1}\right),
\end{align}
where $\mathcal{F}_t$ is the information set up to time $t$, $\zeta_t$ is a vector of time--varying parameters, $\vartheta$ is a vector or static parameters, $\zeta_{t-1},\dots,\zeta_{t-p}$ and $y_{t-1},\dots,y_{t-p}$ are lagged values of the dynamics parameters and the observations, up to order $p\geq 1$, respectively. Finally, the function $h\left(\cdot\right)$ refers to one of the dynamics reported below, while $\mathcal{D}\left(\cdot\right)$ is a specified density function. Moreover, all the model parameters are estimated by maximising the log--likelihood function, see e.g. \cite{francq_zakoian.2011}.
\subsection{ARCH models}
\label{sec:arch_models}
%
\noindent ARCH--type models are flexible and powerful tools for conditional volatility modelling, because they are able to consider the volatility clustering phenomena as well as other established stylised facts. Models belonging to this class, have been principally proposed in order to describe the time--varying nature of the conditional volatility that characterises financial assets. They have also become one of the most used tools for researchers and practitioners dealing with financial market exposure. The simplest conditional volatility dynamics we consider in this paper is the GARCH(p,q) specification introduced by \cite{bollerslev.1986}
\begin{equation}
\sigma_t^2=\omega+\sum_{i=1}^p\alpha_i\varepsilon_{t-i-1}^2+\sum_{j=1}^q\beta_j\sigma_{t-j-1}^2,
\label{eq:garch_pq_spec}
\end{equation}
where $\varepsilon_{t-i}=y_{t-i}-\mu$, for $i=1,2,\dots,p$ and $t=1,2,\dots,T$, $\mu\in\Re$ is the mean of $y_t$, $\omega>0$, $0\leq\alpha_i<1,\forall i=1,2,\dots,p$ and $0\leq\beta_j<1,\forall j=1,2,\dots,q$ with $P\equiv\sum_{i=1}^p\alpha+\sum_{j=1}^q\beta<1$ to preserve weak ergodic stationarity of the conditional variance. Sometimes it is possible to observe high persistence in financial time series volatility, i.e. it is possible to observe series for which $P\approx1$. To account for this scenario the IGARCH(p,q) specification, where the persistence parameters is imposed to be exactly 1, i.e $P=1$, has been proposed by \cite{engle_bollerslev.1986}. Despite their  popularity, the GARCH and the IGARCH specifications are not able to account for returns exhibiting higher volatility after negative shocks than after positive ones as theorised by the \qmo leverage effect\qmcsp of \cite{black.1976}. Consequently, in the financial econometric literature several alternative specifications have been proposed. The EGARCH(p,q) model of \cite{nelson.1991}, for example, assumes that the conditional volatility dynamics follows
\begin{equation}
\log\left(\sigma_t^2\right)=\omega+\sum_{i=1}^p\left[\alpha_i\varpi_{t-i}+\gamma_i\left(\vert\varpi_{t-i}\vert-\mathbb{E}\vert\varpi_{t-i}\vert\right)\right]+\sum_{j=1}^q\beta_j\log\left(\sigma_{t-j}^2\right),
\label{eq:egarch_pq_spec}
\end{equation}
where $\varpi_{t-i}=\frac{\varepsilon_{t-i}}{\sigma_{t-i}}$, for $i=0,1,\dots,p$ and $t=1,2,\dots,T$. The asymmetric response is introduced through the $\gamma_i$ parameters: for $\gamma_i<0$ negative shocks will obviously have a larger impact on future volatility than positive shocks of the same magnitude. Note that no positivity constraints are imposed on the parameters $\alpha_i,\beta_j,\gamma_i$. For the EGARCH(p,q) specification the persistence parameter $P$ is equal to $P=\sum_{j=1}^q\beta_j$.
One of the most flexible models belonging to the ARCH family is the Asymmetric--Power--ARCH(p,q) (APARCH, henceforth) model of \cite{ding_etal.1993} which imposes the following dynamic to the conditional variance
\begin{equation}
\sigma_t^{\delta}=\omega+\sum_{i=1}^p\alpha_i \left(\vert\varepsilon_{t-i}\vert-\gamma_i\varepsilon_{t-i}\right)^{\delta}+\sum_{j=1}^q\beta_j\sigma_{t-j}^{\delta},
\label{eq:aparch_pq_spec}
\end{equation}
where the $\delta$ parameter plays the role of a Box--Cox transformation (see \citealt{box_cox.1964}). To ensure the positiveness of the conditional variance the following parameter restrictions are imposed: $\omega>0$, $\delta\ge0$, $0\le\gamma_i\le0$ for $i=1,\dots,p$ and the usual conditions $\alpha_i\geq0$, and $\beta_j\ge0$, for $i,j=1,2,\dots,\max{\{p,q\}}$. In the APARCH specification the persistence strongly depends upon the distributional assumption made on $y_t$, i.e.
\begin{equation}
P=\sum_{i=1}^p\alpha_i\kappa_i + \sum_{j=1}^q\beta_j,
\end{equation}
where $\kappa_i=\mathbb{E}\left[\vert\varpi_t\vert-\gamma_i\varpi_t\right]^\delta$, for $i=1,\dots,p.$
The APARCH specification results in a very flexible model that nests several of the most popular univariate ARCH parameterisations, such as
\begin{itemize}
\item[\textit{(i)}] the GARCH(p,q) of \citet{bollerslev.1986}, for $\delta=0$ and $\gamma_i=0$, for $i=1,2,\dots,p$;
\item[\textit{(ii)}] the Absolute--Value--GARCH (AVARCH, henceforth) specification, for $\delta=1$ and $\gamma_i=0$ for $i=1,2,\dots,p$, proposed by \citet{taylor.1986} and \citet{schwert.1990} to mitigates the influence of large, in an absolute sense, shocks with respect to the traditional GARCH specification;
\item[\textit{(iii)}] the GJR--GARCH model (GJRGARCH, henceforth) of \citet{glosten_etal.1993}, for $\delta=2$ and $0\le\gamma_i\le1$ for $i=1,2,\dots,p$;
\item[\textit{(iv)}] the Threshold GARCH (TGARCH, henceforth) of \citet{zakoian.1994}, for $\delta=1$, which allows different reactions of the volatility to different signs of the lagged errors;
\item[\textit{(v)}] the Nonlinear GARCH (NGARCH, henceforth) of \citet{higgins_bera.1992}, for $\gamma_i=0$ for $i=1,2,\dots,p$ and $\beta_j=0$ for $j=1,2,\dots,q$.
\end{itemize}
Another interesting specification is the Component--GARCH(p,q) of \citet{engle_lee.1993} (CGARCH, henceforth) which decomposes the conditional variance into a permanent and transitory component in a straightforward way
\begin{align}
\sigma_t^2&=\xi_t+\sum_{i=1}^p\alpha_i\left( \varepsilon_{t-i}^2-\xi_{t-i}\right)+\sum_{j=1}^q\beta_j\left(\sigma_{t-j}^2-\xi_{t-j}\right)\nonumber\\
\xi_t&=\omega+\rho \xi_{t-1}+\eta\left(\varepsilon_{t-1}^2-\sigma_{t-1}^2\right),
\label{eq:csgarch_pq_spec}
\end{align}
where in order to ensure the stationarity of the process we impose $\sum_{i=1}^p\alpha_i + \sum_{j=1}^q\beta_j<1$ and the additional condition that $\rho<1$. Further parameters restrictions for the positiveness of the conditional variance are given in \cite{engle_lee.1993}. This solution is usually employed because it permits to investigate the long and short--run movements of volatility.
The considered conditional volatility models are a minimal part of the huge number of specifications available in the financial econometric literature. We chose these models because of their heterogeneity, since each of them focuses on a different kind of stylised fact. Moreover, even if they could seem very similar, the way in which they account for the stylised fact changes. For a very extensive and up to date survey on GARCH models we will refer the reader to the works of \cite{bollerslev.2008}, \cite{terasvirta.2009}, \cite{bauwens_etal.2006}, \cite{silvennoinen_terasvirta.2009} and the recent book of \cite{francq_zakoian.2011}.
%
\subsection{GAS models}
\label{sec:gas_models}
%
\noindent The GAS framework recently introduced by \cite{creal_etal.2013} and \cite{harvey.2013} is gaining lots of consideration by econometricians in many field of time series analysis. Under the \cite{cox.1981} classification the GAS models can be considered as a class of observation driven models, with the usual consequences of having a closed form for the likelihood and ease of evaluation. The key feature of GAS models is that the predictive score of the conditional density is used as forcing variable into the updating equation of a time--varying parameter. Two main reasons for adopting this updating procedure has been given in literature. \cite{harvey.2013}, for example, argues that the GAS specification can be seen as an approximation to a filter for a models driven by a stochastic latent parameter that is, by definition, \qmo unobservable\qmc. \cite{creal_etal.2013} instead consider the conditional score as a steepest ascent direction for improving the model's local fit given the current parameter position, as it usually happens into a Newton--Raphson algorithm. Moreover, the flexibility of the GAS framework make this class of models nested with an huge amount of famous econometrics models such as, for example, some of the ARCH--type models of \cite{engle.1982} and \cite{bollerslev.1986} for volatility modelling, and also the MEM, ACD and ACI models of \cite{engle.2002b}, \cite{engle_gallo.2006}, \cite{engle_russell.1998} and \cite{russell.1999}, respectively. Finally, one of the practical implications of using this framework in order to update the time--varying parameters is that it avoids the problem of using a non--adequate forcing variable when the choice of it is not so obvious. In fact, it may be argued that, the use of the conditional score vector, allows for the dynamic parameter to be updated considering all the information coming from the entire distribution as usually happen into a state space framework. On the contrary, many others observation driven models, such as ARCH--type models, only use the expected value of the conditional distribution. GAS model applications range in several interesting areas, such as, risk measure and dependence modelling, \cite{bernardi_catania.2015}, \cite{lucas_zhang.2014} and \cite{salvatierra_patton.2014}, volatility modelling, \cite{harvey_sucarrat.2014}, and in a nonlinear autoregressive setting, by \cite{koopman_etal.2015} and \cite{dellemonache_petrella.2014}.
\newline
\indent Formally, let us consider the general model specification in Equations \eqref{eq:general_model_spec_meas}--\eqref{eq:general_model_spec_trans}, where, as before, $\mathcal{D}\left(\cdot\right)$ denotes a probability density, $\zeta_t$ is a set of time--varying parameters and $\vartheta$ is a vector of time--independent parameters. For example, $\mathcal{D}\left(\cdot\right)$ may be a Student--t distribution with fixed degree of freedom ($\lambda=\nu$) and time--varying volatility ($\zeta_t=\sigma_t$). Then, the updating equation for the time--varying parameters according to the GAS  framework is
\begin{align}
\zeta_{t+1}&=\omega_\zeta + \alpha_\zeta s_t + \beta_\zeta\zeta_t\nonumber\\
s_t&=S_t\left(\zeta_t\mid\vartheta\right)\nabla_t\left(\zeta_t\mid\vartheta\right),\nonumber
\end{align}
where $\nabla_t\left(\zeta_t\mid\vartheta\right)$ is the conditional score of the pdf $\mathcal{D}\left(\cdot\right)$, evaluated at $\zeta_t$
\begin{equation}
\nabla_t\left(\zeta_t\mid\vartheta\right)=\frac{\partial\ln \mathcal{D}\left(\zeta_t\mid\vartheta\right)}{\partial \zeta_t},\nonumber
\end{equation}
and $S_t\left(\zeta_t\mid\vartheta\right)$ is a positive definite, possible parameter--dependent scaling matrix. A convenient choice for the scaling matrix $S_t\left(\zeta_t\mid\vartheta\right)$ is usually given by the Fisher information matrix
\begin{equation}
S_t\left(\zeta_t\mid\vartheta\right)=\left[\mathcal{I}\left(\zeta_t\mid\vartheta\right)\right]^{-a},
\end{equation}
where $\mathcal{I}\left(\zeta_t\mid\vartheta\right)$ can be written as:
\begin{eqnarray}
\mathcal{I}\left(\zeta_t\mid\vartheta\right)=-\mathbb{E}_{t-1}\left[\frac{\partial^2\ln \mathcal{D}\left(\zeta_t\mid\vartheta\right)}{\partial \zeta_t\partial \zeta_t^\top}\right]= \mathbb{E}_{t-1}\left[\nabla_t\left(\zeta_t\mid\vartheta\right)\times \nabla_t\left(\zeta_t\mid\vartheta\right)^\top\right],
\end{eqnarray}
and $a$ is usually set equal to $\{0,\frac{1}{2},1\}$. Note that for $a=0$ the scaling matrix $S_t\left(\zeta_t\mid\vartheta\right)$ coincides with the identity matrix. \cite{creal_etal.2013} suggest to use the inverse Fisher information matrix ($\alpha=1$) or its (pseudo)--inverse square root ($\alpha=\frac{1}{2}$) in order to scale the conditional score for a quantity that accounts for its curvature (variance). In our empirical tests we find this way much more efficient than using an identity scaling matrix. However, sometimes the Fisher information matrix is not available in closed form, and simulation and numerical evaluation techniques should be used to approximate the hessian matrix.\newline
\indent Another interest property of the GAS framework is that it adapts quite naturally to various reparameterisations of the problem in hand. This aspect is quite useful when the natural parameters' space is constrained into a subset of the real line $\Re$, and therefore a mapping function $\vartheta\left(\cdot\right)$ between this and the real line, becomes necessary. For example, let us define $\Im\in\Re$ as the natural parameters' space and $\vartheta : \Re\to\Im$ an absolutely continuous deterministic invertible mapping function that maps the real line into the natural parameter space $\Im$. In general we consider the modified logistic function defined by
\begin{equation}
m_{\left(\mathrm{L},\mathrm{U}\right)}\left(x\right)=\mathrm{L} + \frac{\left(\mathrm{U}-\mathrm{L}\right)}{1+e^{-x}},
\end{equation}
which maps $\Re$ into the interval $\left(\mathrm{L},\mathrm{U}\right)$. Moreover, let us define $\tilde{\zeta}_t=m^{-1}\left(\zeta_t\right)$ as the unmapped version of the parameter $\zeta_t$, then the GAS model suited for the new time--varying parameter $\tilde \zeta_t$ with $a=1$ is defined as
\begin{equation}
\tilde{\zeta}_{t+1} = \omega_\zeta + \alpha_\zeta \tilde s_t  + \beta_\zeta \tilde{\zeta}_t,
\end{equation}
where $\tilde s_t=\dot{m}_t s_t$, and $\dot{m}_t=\frac{\partial m\left(\tilde\zeta_t\right)}{\partial \tilde\zeta_t}$. For other possible choices of $a$ we refer to \cite{creal_etal.2013}.\newline
%
%
\indent In the empirical application, we will consider the GAS specification for the parameters of the Gaussian and Student--t distributions. Formally, the Gaussian GAS model for the parameters $\zeta_t^{\sN}=\left(\mu_t,\sigma_t^2\right)^\prime$ is given by:
\begin{equation}
y_t\mid\left(\mathcal{F}_{t-1},\zeta_t^\sN,\vartheta\right) \sim\mathcal{N}\left(\mu_t,\sigma_t^2\right),\qquad t=1,2,\dots,T,\nonumber
\end{equation}
with
\begin{align}
 \begin{pmatrix}
 \mu_{t+1}\\
  \tilde\sigma_{t+1}^2
 \end{pmatrix} =
  \begin{pmatrix}
 \omega_{\mu}\\
  \omega_{\sigma^2}
 \end{pmatrix} +
   \begin{pmatrix}
 \alpha_{\mu} & 0 \\
  0 & \alpha_{\sigma^2}
 \end{pmatrix} \tilde{s}_t  +
    \begin{pmatrix}
 \beta_{\mu} & 0 \\
  0 & \beta_{\sigma^2}
 \end{pmatrix}
 \begin{pmatrix}
 \mu_{t}\\
  \tilde\sigma_{t}^2
 \end{pmatrix},\nonumber
\end{align}
where $\tilde\sigma_t^2=\log\left(\sigma_t^2\right)$ and borrowing the previous notation
\begin{align}
\dot{m}_t=
\begin{bmatrix}
1\\
1/\tilde\sigma^2_t
\end{bmatrix},\quad\mathcal{I}_t=\begin{bmatrix}
 \frac{1}{\sigma_t^2} & 0 \\
  0 & \frac{1}{2\sigma_t^4}
 \end{bmatrix},\quad\nabla_t=\begin{bmatrix}
 \frac{\left(y_t - \mu_{t}\right)}{\sigma_t^2} \\
  -\frac{1}{2\sigma_t^2}\left(1-\frac{\left(y_t - \mu_{t}\right)^2}{\sigma_t^2}\right)
 \end{bmatrix}.\nonumber
\end{align}
The Student--t GAS model with time varying location, scale and shape parameters $\zeta_t^\sT=\left(\mu_t,\phi_t^2,\nu_t\right)^\prime$ is given by:
\begin{equation}
y_t\mid\left(\mathcal{F}_{t-1},\zeta_t^\sT,\vartheta\right)\sim\mathcal{T}\left(\mu_t,\phi_t^2,\nu_t\right),\qquad t=1,2,\dots,T,\nonumber\\
\end{equation}
with
\begin{align}
 \begin{pmatrix}
 \mu_{t+1}\\
  \tilde\phi_{t+1}^2\\
  \tilde\nu_{t+1}
 \end{pmatrix} =
  \begin{pmatrix}
 \omega_{\mu}\\
  \omega_{\phi}\\
  \omega_{\nu}\\
 \end{pmatrix} +
   \begin{pmatrix}
 \alpha_{\mu} & 0 & 0\\
  0 & \alpha_{\phi} & 0\\
  0 & 0 & \alpha_{\nu} \\
 \end{pmatrix} \tilde{s}_t  +
    \begin{pmatrix}
 \beta_{\mu} & 0 & 0\\
  0 & \beta_{\phi} & 0\\
  0 & 0 & \beta_{\nu}\\
 \end{pmatrix}
 \begin{pmatrix}
 \mu_{t}\\
  \tilde\phi_{t}^2\\
  \tilde\nu_{t}
 \end{pmatrix},\nonumber
\end{align}
where $\tilde\phi_t^2=\log\left(\phi_t^2\right)$, $\tilde\nu_t=\log\left(\nu_t\right)$, and
\begin{align}
\dot{m}_t &=\begin{pmatrix}
 1\\
 1/\tilde\phi^2_t\\
  1/\tilde\nu_t
 \end{pmatrix},\nonumber\\
 \mathcal{I}_t&=\begin{pmatrix}
 \frac{\nu_t+1}{\phi_t^2\left(\nu_t+3\right)} & 0  & 0\\
  0 & \frac{\nu_t}{2\phi_t^4\left(\nu_t+3\right)} & \frac{1}{2\phi^2\left(\nu_t+3\right)\left(\nu_t+1\right)} \\
  0 & \frac{1}{2\phi^2\left(\nu_t+3\right)\left(\nu_t+1\right)} & \frac{h_1\left(\nu_t\right)}{2}
 \end{pmatrix},\nonumber\\
 \nabla_t&=\begin{pmatrix}
\frac{\left(\nu_t+1\right)\left(r_t-\mu_t\right)}{2\phi_t^2 + \left(r_t-\mu_t\right)^2}\nonumber\\
-\frac{1}{\phi_t} + \frac{\left(\nu_t+1\right)\left(r_t-\mu_t\right)}{\nu_t\phi_t^3+\phi_t\left(r_t-\mu_t\right)}\nonumber\\
h_2\left(r_t,\mu_t,\phi_t^2,\nu_t\right)\nonumber
 \end{pmatrix},
\end{align}
and
\begin{align}
h_1\left(\nu_t\right)&=-\frac{1}{2}\psi'\left(\frac{\nu_t+1}{2}\right) + \frac{1}{2}\psi'\left(\frac{\nu_t}{2}\right) - \frac{\nu_t+5}{\nu_t\left(\nu_t+3\right)\left(\nu_t+1\right)}\nonumber\\
h_2\left(r_t,\mu_t,\phi_t^2,\nu_t\right)&=\frac{1}{2}\psi\left(\frac{\nu_t+1}{2}\right) - \frac{1}{2}\psi\left(\frac{\nu_t}{2}\right) - \frac{\pi}{2\nu_t} - \nonumber\\
&\qquad\qquad\frac{1}{2}\log\left(1+\frac{\left(r_t-\mu_t\right)^2}{\nu_t\phi_t^2}\right)
+ \frac{\left(\nu_t+1\right)\left(r_t-\mu_t\right)^2}{2\nu_t\left(\phi^2\nu_t+\left(r_t-\mu_t\right)^2\right)},\nonumber
\end{align}
where $\psi\left(x\right)$ and $\psi'\left(x\right)$ are the digamma and the trigamma functions, respectively.
Recently, an OxMetrics package providing functions to estimate various specifications of GAS models has been developed by \cite{andres.2014}.
%
\subsection{Dynamic quantile models}
\label{sec:caviar_models}
%
\noindent The CAViaR models of \cite{engle_manganelli.2004}, extends the standard quantile regression model introduced by \cite{koenker_bassett.1978} and belongs to the family of dynamic quantile autoregressive models proposed by \cite{koenker_xiao.2006}. Formally, let $f_t\left(\bbeta_\tau\right)\equiv f_t \left(\bx_{t-1},\bbeta_\tau\right)$ denote the $\tau$--th level conditional quantile at time $t$ of the observed variable $y_t$ conditional to the information available at time $t-1$, i.e. $\bx_{t-1}$ and the vector of unknown parameters $\bbeta_\tau$, the generic CAViaR specification for the $\tau$--th quantile can be written as:
\begin{align}
\label{eq:caviar_meas}
y_t&=f_t \left(\bbeta_\tau\right)+\epsilon_t,\qquad\qquad t=1,2,\dots,T,\\
\label{eq:caviar_trans}
f_t \left(\bbeta_\tau\right)&=\theta^{\tau}+\sum_{i=1}^p\theta^{\tau}_i f_{t-i}\left(\bbeta_\tau\right)+\sum_{j=1}^q\phi^{\tau}_{j}\ell\left(\bx_{t-j}\right),
\end{align}
where $\btheta=\left(\theta_0,\theta_1,\dots,\theta_p\right)\in \Re^{p+1}$ collects the autoregressive parameters, while $\bphi=\left(\phi_1,\phi_2,\dots,\phi_q\right)\in \Re^{q}$, $\ell\left(\cdot\right)$ is the function linking the lagged exogenous information or past returns $y_{1:t-1}=\left(y_1,y_2,\dots,y_{t-1}\right)$ to the current conditional quantile and the error term $\epsilon_t$ in the measurement Equation \eqref{eq:caviar_meas} is such that its $\tau$--th level quantile is equal to zero, i.e. $q_\tau\left(\epsilon_t\mid \mathbf{x}_{1:t-1}\right)=0$ in order to ensure that $f_t\left(\bbeta_\tau\right)$ is the $\tau$--th level conditional quantile of the observed variable $y_t$ given the observations up to time $t-1$.
As noted by \cite{engle_manganelli.2004}, the smooth evolution of the quantile over time is guaranteed by the autoregressive terms denoted by $\beta_i f_{t-i}\left(\bbeta\right)$, $i=1,2,\dots,p$, while the function $\ell\left(\cdot\right)$ can be interpreted as the News Impact Curve (NIC) introduced by \cite{engle_ng.1993} for ARCH--type models.\newline
\indent The estimation procedure of the $\tau$--th regression quantile in the frequentist approach is based on the minimisation of the following loss function
\begin{equation}
\min_\beta \sum_t\rho_{\tau}\left( y_{t}-f_t \left(\bbeta_\tau\right)\right),
\label{eq:quantile_loss_fuction}
\end{equation}
with $\rho_{\tau} \left(u\right)=u\left(\tau-\bbone\left(u<0\right)\right)$. Alternatively, the Asymmetric Laplace distribution (ALD) can be used as misspecified likelihood function to perform maximum likelihood inference as suggested by \cite{yu_moyeed.2001} from a Bayesian perspective. They also prove that, under improper prior for the regression parameters $\bbeta_\tau$, the Baysian Maximum a Posteriori estimate (MaP) coincides with the solution of the minimisation problem in equation \eqref{eq:quantile_loss_fuction}.
Some examples of specifications of the CAViaR dynamic in equation \eqref{eq:caviar_trans} have been introduced in the seminal paper of \cite{engle_manganelli.2004}:
\begin{itemize}
\item[\textit{(i)}] symmetric absolute value:
\begin{equation}
f_t\left(\bbeta\right)=\beta_1+\beta_2 f_{t-1}\left(\bbeta\right)+\beta_3\vert y_{t-1}\vert,
\end{equation}
\item[\textit{(ii)}] asymmetric slope
\begin{equation}
f_t\left(\bbeta\right)=\beta_1+\beta_2 f_{t-1}\left(\bbeta\right)+\beta_3y_{t-1}\bbone_{\left[0,+\infty\right)}\left(y_{t-1}\right)
+\beta_4y_{t-1}\bbone_{\left(-\infty,0\right)}\left(y_{t-1}\right),
\end{equation}
\item[\textit{(iii)}] indirect GARCH(1,1)
\begin{equation}
f_t\left(\bbeta\right)=\left[\beta_1+\beta_2 f_{t-1}^2\left(\bbeta\right)+\beta_3y_{t-1}^2\right]^{\frac{1}{2}},
\end{equation}
\item[\textit{(iv)}] adaptive
\begin{equation}
f_t\left(\bbeta\right)=f_{t-1}\left(\bbeta\right)+\frac{\beta_1}{1+\exp\left\{G\left(y_{t-1}-f_{t-1}\left(\bbeta\right)\right)\right\}},
\end{equation}
where $G\in\Re^+$ is a positive constant.
\end{itemize}
The different specifications introduced by the original paper of \cite{engle_manganelli.2004} can be estimated using the code available at the first author's web page: \url{http://www.simonemanganelli.org/Simone/Research.html}.
%
\section{The MCS package}
\label{sec:mcs_package}
%
\noindent As described in Section \ref{sec:mcs_procedure}, the MCS procedure is used to compare different models under an user defined loss function. The loss function measures the \qmo performance\qmcsp of the competing models at a each time point $t=1,2,\dots$ in the evaluating period. Suppose now to compare $m$ alternative models over the evaluating period of length $n$, then the loss function defined in Equation \ref{eq:loss} delivers a loss matrix of dimension $\left(m\times n\right)$ containing, for each time $t=1,2,\dots$, the losses associated to each competing model. The \textsf{R} function {\tt MCSprocedure()} is then used to construct the set of superior models outlined in Section \ref{sec:mcs_procedure}.
%
\subsection{Comparing models using the MCS routine}
\label{sec:mcs_garch_R}
%
\noindent The MCS procedure can be used to compare models under various aspects. For example, it can be used to assess the models' ability to predict future volatility levels or future returns, conditional to actual and past information. The object \qmo{\tt Loss}\qmcsp represents the main input of the implemented {\tt MCSprocedure()} function. It consists of a matrix of dimension $\left(m\times n\right)$, where $m$ is the number of competing models in the initial set ${\rm M}^0$ and $n$ is the number of observations in the evaluation period. In the next section, we describe some alternative loss functions specifications which are particularly suitable for volatility forecast assessment as well as to forecast future observations.
%
%
\subsection{Loss functions}
\label{sec:loss_functions}
%
\noindent As previously discussed, the MCS procedure is able to discriminate models under a user defined loss function. The choice of the loss function is somewhat arbitrary, and it crucially depends on the nature of the competing models and the scope of their usage. For more considerations about the choice of the loss function for model comparison purposes, we refer to \cite{hansen_lunde.2005}, \cite{bollerslev_etal.1994}, \cite{diebold_lopez.1996} and \cite{lopez.2001}. In what follows, we report the loss functions available within the \texttt{MCS} package. However, since the \texttt{MCSprocedure()} function accepts as input a pre--defined loss matrix, named \qmo\texttt{Loss}\qmc, the user is free to define and use its own loss function.
The following loss functions are freely available within the \texttt{MCS} package:
\begin{itemize}
\item[1.] the \texttt{LossVaR()} that can be used to check the performances associated to VaR, or, more generally, quantile forecasts;
\item[2.] the \texttt{LossVol()} for volatility forecasts assessment;
\item[3.] the \texttt{LossLevel()} that can be used instead for level forecasts, as the punctual mean forecasts of a regression model.
\end{itemize}
These loss functions accept as inputs common and function specific arguments. The common arguments are \qmo\texttt{realized}\qmc, that consists of a vector of realised observations, i.e. the ones that a model hopes to accurately forecast or describe, and \qmo\texttt{evaluated}\qmc, which is a vector or a matrix of models output. It is worth noting that we decided to call the second argument of those functions \qmo \texttt{evaluated}\qmcsp instead of \qmo \texttt{forecasted}\qmcsp since the MCS procedure is more general than a simple procedure for forecasts evaluation. In fact, as reported by \cite{hansen_etal.2011}, the MCS procedure also adapts to in sample studies. The third argument, \qmo\texttt{which}\qmcsp, instead is function specific. The available choices for the common and function specific arguments are reported below.
\begin{itemize}
\item[1.] Concerning the \texttt{LossVaR()} function, the only available argument is \texttt{which = "asymmetricLoss"}. This coincides with the asymmetric VaR loss function of \cite{gonzalez_etal.2004} defined in equation \ref{eq:loss_function_asymmetric} which is used to assess quantile--based risk measures, such as the VaR, because it penalises more heavily observations below the $\tau$--th level quantile, i.e. $y_t<\mathrm{VaR}_t^{\tau}$. 
Further arguments of the \texttt{LossVaR()} function are: \qmo\texttt{tau}\qmc, which represents the VaR confidence level, and \texttt{type=\{"normal","differentiable"\}}. The \qmo\texttt{type}\qmcsp argument allows to discriminate between the normal and the differentiable versions of the loss function. The choice \qmo normal\qmcsp specifies the loss function of \cite{gonzalez_etal.2004} defined in equation \ref{eq:loss_function_asymmetric}, while the option \qmo differentiable\qmcsp  considers the following loss function
\begin{equation}
\ell\left(r_{t},\mathrm{VaR}_{t}^\tau\right)=\left(\tau-m_{\delta}\left(r_{t},\mathrm{VaR}_{t}^\tau\right)\right)\left(r_{t}-\mathrm{VaR}_{t}^\tau\right),
\end{equation}
where $m_\delta\left(a,b\right)=\left[1+\exp{\{\delta\left(a-b\right)\}}\right]^{-1}$. Note that the parameter $\delta$, controlling for the function smoothness, it is fixed to the default value of $25$, but different value can be specified by the user.
\item[2.] Concerning the \texttt{LossVol()}, we implemented the functions reported in \cite{hansen_lunde.2005}. Note that for this kind of loss functions the \texttt{realized} and \texttt{evaluated} arguments should be some realised volatility measures $\tilde\sigma_{t+1}$ and the punctual volatility forecasts $\hat\sigma_{t+1}$. In this context, we use the term volatility as for the standard deviation. The implemented loss functions are:
\begin{itemize}
\item[2.1] $\mathrm{SE}_{1,t+1}=\left(\tilde\sigma_{t+1}-\hat\sigma_{t+1}\right)^2$, by setting \texttt{which = "SE1"};
\item[2.2] $\mathrm{SE}_{2,t+1}=\left(\tilde\sigma_{t+1}^2-\hat\sigma_{t+1}^2\right)^2$, by setting \texttt{which = "SE2"};
\item[2.3] $\mathrm{QLIKE}_{t+1}=\log{\left(\hat\sigma_{t+1}^2\right)+\tilde\sigma_{t+1}^2 \hat\sigma_{t+1}^{-2}}$, by setting \texttt{which = "QLIKE"};
\item[2.4] $\mathrm{R}^2\mathrm{LOG}_{t+1}=\left[\log\left(\tilde\sigma_{t+1}^2 \hat\sigma_{t+1}^{-2}\right)\right]^2$, by setting \texttt{which = "R2LOG"};
\item[2.5] $\mathrm{AE}_{1,t+1}=\vert \tilde\sigma_{t+1} - \hat\sigma_{t+1}\vert$, by setting \texttt{which = "AE1"};
 \item[2.6] $\mathrm{AE}_{2,t+1}=\vert \tilde\sigma_{t+1}^2 - \hat\sigma_{t+1}^2\vert$, by setting \texttt{which = "AE2"}.
\end{itemize}
\item[3.] Concerning the \texttt{LossLevel()} function, the only available argument is \texttt{which = \{"SE", "AE"\}}. The two options coincide with the squared error and the absolute error, respectively.
\end{itemize}
%
\subsection{Constructing the SSM}
\label{sec:ssm_garch_R}
%
\noindent The function {\tt MCSprocedure()} is the main routine for the MCS procedure. It returns a \textsf{R} \qmo S4\qmcsp object of the class \qmo{\tt SSM}\qmc, which has several arguments we now briefly describe here. The main inputs of the function {\tt MCSprocedure()} are:
\begin{itemize}
\item[-] \texttt{"Loss"}, a matrix or something coercible to that (using the as.matrix() function), containing the loss series for each competing model;
\item[-] \texttt{"alpha"}, a positive scalar in $\left(0,1\right)$ indicating the confidence level of the MCS tests;
\item[-] \texttt{"B"}, an integer indicating the number of bootstrapped samples used to construct the statistic test;
\item[-] \texttt{"cluster"}, a cluster object created by calling \texttt{makeCluster()} from the \textsf{R} \texttt{parallel} package. The default option for  \texttt{"cluster"} is set to \texttt{NULL}, otherwise the user supplied cluster object is employed in the parallel processing routine.
\item[-] \texttt{"statistic"}, defines the test--statistic that is used to test the EPA. Possible choices are \qmo Tmax\qmcsp and \qmo TR\qmc, which coincide with $\mathrm{T}_{\max ,{\rm M}}$ and $\mathrm{T}_{\rm R,M}$ statistics defined in Section \ref{sec:mcs_procedure}.
\end{itemize}
%
%
\section{Application}
\label{sec:application}
%
\noindent In this empirical study, a panel of four major worldwide stock markets indexes is considered. The four daily stock price indices includes the Asia/Pacific 600 (SXP1E), the North America 600 (SXA1E), the Europe 600 (SXXP) and the Global 1800 (SXW1E). The data are freely available and can be downloaded from the STOXX website \url{http://www.stoxx.com/indices/types/benchmark.html}. The data were obtained over a 23--years time period, from December 31, 1991 to July 24, 2014, comprising a total of 5874 observations. For each market, the returns are calculated as the logarithmic difference of the daily index level multiplied by 100
\begin{equation}
y_t=\left(\log\left(p_t\right)-\log\left(p_{t-1}\right)\right)\times 100,\nonumber
\end{equation}
where $p_t$ is the closing index value on day $t$.
To examine the performance of the models to predict extreme VaR levels, the complete dataset of daily returns is divided into two samples: an in--sample period from January 1, 1992 to  October 6, 2006, for a total of 3814 observations, and a forecast or validation period, containing the remaining 2000 observations: from October 9, 2006 to July 24, 2014. A rolling window approach is then used to produce 1--day ahead forecasts of the 5\% VaR thresholds, $\mathrm{VaR}_{t+1}^{0.05}$, for $t=1,2,\dots,2000$ of the considered series in the forecast sample.
Table \ref{tab:Index_data_summary_stat} reports some descriptive statistics for the in sample as well as the out of sample period. As expected, we found evidence of departure from normality, mainly because all the series appear to be leptokurtic and skewed. Moreover, the \cite{jarque_bera.1980} statistic test strongly rejects the null hypothesis of normality for all the considered series. It is interesting to note that, the departure from normality, is stronger for the out of sample returns. As widely discussed by \cite{shiller_2012}, this empirical evidence can be considered as an effect of the recent GFC of 2007--2008 that affected the overall economy. Furthermore, the unconditional distribution of each return's series, in the out of sample period, is negatively skewed and shows higher standard deviation as well as higher kurtosis. The 5\% unconditional quantile, which represents the VaR at $\tau=5\%$ under the iid assumption of the returns' conditional distribution, has been moved further to the left tail, in the second part of the sample. The changes in the behaviour of the considered panel of returns suggest that the SSM would contain those models that are more flexible to describe the impact of the GFC to the returns' conditional distributions during the validation period.\newline
\indent As previously said, our empirical application focuses on the ability to forecast the Value--at--Risk of several competing models, at a given confidence level $\tau$. To apply the MCS procedure, we forecast the VaR at $\tau=5\%$ using the models described in Section \ref{sec:model_specifications} estimated on each of the four indexes. More precisely, we consider eight different GARCH(1,1) specifications (GARCH, EGARCH, APARCH, AVARCH, GJRGARCH, TGARCH, NGARCH, CGARCH) with Gaussian and Student--t innovations detailed in Section \ref{sec:arch_models}, the GAS--$\mathcal{N}$ and the GAS--$\mathcal{T}$ models described in section \ref{sec:gas_models}, and four CaViaR model specifications reported in Section \ref{sec:caviar_models} comprising a total of $22$ models. Estimated coefficients for each model are not reported to save space, but they are available upon request to the second author. For the GARCH and the GAS models, VaR estimates are performed by inverting the corresponding conditional cumulative density function, while the CAViaR specification reports directly the quantile estimates. The MCS procedure of \citet{hansen_etal.2011} described in Section \ref{sec:mcs_procedure} is then applied to obtain the set of models with superior predictive ability in term of the supplied VaR forecasts.\newline
%
\begin{table}[!t]
\captionsetup{font={small}, labelfont=sc}
\begin{small}
\resizebox{1.0\columnwidth}{!}{%
\centering
\smallskip
\begin{tabular}{lccccccccc}\\
\toprule
Index & Min & Max & Mean & Std. Dev. & Skewness & Kurtosis & 5\% Str. Lev. & JB \\
\hline
\multicolumn{9}{l}{\textit{In--sample, from 02/01/1992 to 06/10/2006}}\\
SXA1E & -8.05 & 7.79 & 0.03 & 1.28 & -0.07 & 5.82 & -2.08 & 1269.17 \\
SXP1E & -5.80 & 9.71 & 0.01 & 1.29 & 0.10 & 5.90 & -1.99 & 1341.19 \\
SXW1E & -5.54 & 5.02 & 0.03 & 0.99 & -0.07 & 5.57 & -1.60 & 1055.46 \\
SXXP & -6.41 & 5.64 & 0.03 & 1.05 & -0.27 & 6.82 & -1.68 & 2376.35 \\
\hline
Index & Min & Max & Mean & Std. Dev. & Skewness & Kurtosis & 1\% Str. Lev. & JB \\
\hline
\multicolumn{9}{l}{\textit{Out--of--sample, from 09/10/2006 to 24/06/2014}}\\
SXA1E & -9.18 & 9.96 & 0.02 & 1.38 & -0.28 & 11.32 & -2.13 & 5812.50 \\
SXP1E & -7.73 & 9.42 & 0.00 & 1.28 & -0.33 & 8.72 & -2.02 & 2768.92 \\
SXW1E & -6.81 & 8.37 & 0.01 & 1.05 & -0.28 & 10.43 & -1.62 & 4641.35 \\
SXXP & -7.93 & 9.41 & 0.00 & 1.33 & -0.11 & 9.61 & -2.06 & 3655.56 \\
\bottomrule
\end{tabular}}
\caption{\footnotesize{Summary statistics of the panel of international indexes, for the in sample and out of sample period. The seventh column, denoted by ``5\% Str. Lev.'' is the 5\% empirical quantile of the returns distribution, while the eight column, denoted by ``JB'' is the value of the Jarque-Ber\'a test-statistics}}
\label{tab:Index_data_summary_stat}
\end{small}
%
\end{table}

\begin{table}[!ht]
\centering
\resizebox{0.9\columnwidth}{!}{%
\begin{tabular}{lccccccc}
\toprule
Model & $\mathrm{Rank}_{\mathrm{R,M}}$ & $t_{ij}$	& $\mathrm{p-value}_{\mathrm{R,M}}$	&  $\mathrm{Rank}_{\mathrm{max,M}}$  & $t_{i\cdot}$ & $\mathrm{p-value}_{\mathrm{max,M}}$	& Loss$\times10^3$ \\
\hline
\multicolumn{8}{l}{\textit{SXA1E: 5 eliminations}}\\
GJRGARCH--$ \mathcal{T}$ & 1 & -1.73  &  1.00  & 1 & -0.45  &  1.00  &  38.90  \\
GJRGARCH--$ \mathcal{N}$ & 2 & -1.28  &  1.00  & 2 &  0.45  &  1.00  &  39.20  \\
AVGARCH--$ \mathcal{N}$ & 3 & -0.99  &  1.00  & 3 &  0.67  &  1.00  &  39.40  \\
APARCH--$ \mathcal{T}$ & 4 & -0.96  &  1.00  & 4 &  0.69  &  1.00  &  39.40  \\
APARCH--$ \mathcal{N}$ & 5 & -0.52  &  1.00  & 5 &  0.98  &  0.98  &  39.70  \\
EGARCH--$ \mathcal{T}$ & 6 & -0.20  &  1.00  & 6 &  1.19  &  0.90  &  39.80  \\
TGARCH--$ \mathcal{T}$ & 7 &  0.01  &  1.00  & 7 &  1.30  &  0.82  &  39.90  \\
NGARCH--$ \mathcal{T}$ & 8 &  0.26  &  1.00  & 8 &  1.42  &  0.70  &  40.10  \\
GARCH--$ \mathcal{T}$ & 9 &  0.31  &  1.00  & 9 &  1.45  &  0.67  &  40.10  \\
GAS--$ \mathcal{T}$ & 10 &  0.39  &  1.00  & 10 &  1.49  &  0.62  &  40.10  \\
NGARCH--$ \mathcal{N}$ & 11 &  0.70  &  0.99  & 11 &  1.60  &  0.48  &  40.30  \\
AVGARCH--$ \mathcal{T}$ & 12 &  0.72  &  0.98  & 13 &  1.68  &  0.38  &  40.30  \\
GARCH--$ \mathcal{N}$ & 13 &  0.80  &  0.96  & 12 &  1.66  &  0.40  &  40.40  \\
TGARCH--$ \mathcal{N}$ & 14 &  1.26  &  0.59  & 14 &  1.90  &  0.11  &  40.70  \\
CGARCH--$ \mathcal{T}$ & 15 &  1.39  &  0.42  & 15 &  1.98  &  0.05  &  40.70  \\
\hline
\multicolumn{8}{l}{\textit{SXP1E: 2 eliminations}}\\
GAS--$ \mathcal{T}$ & 1 & -1.94  &  1.00  & 1 & -0.79  &  1.00  &  38.80  \\
GJRGARCH--$ \mathcal{N}$ & 2 & -1.19  &  1.00  & 2 &  0.78  &  1.00  &  39.50  \\
AVGARCH--$ \mathcal{T}$ & 3 & -1.05  &  1.00  & 3 &  0.84  &  1.00  &  39.50  \\
GAS--$ \mathcal{N}$ & 4 & -0.83  &  1.00  & 4 &  1.01  &  0.99  &  39.70  \\
APARCH--$ \mathcal{N}$ & 5 & -0.60  &  1.00  & 5 &  1.15  &  0.96  &  39.80  \\
AVGARCH--$ \mathcal{N}$ & 6 & -0.50  &  1.00  & 6 &  1.16  &  0.95  &  39.90  \\
GJRGARCH--$ \mathcal{T}$ & 7 & -0.50  &  1.00  & 7 &  1.23  &  0.92  &  39.90  \\
APARCH--$ \mathcal{T}$ & 8 & -0.13  &  1.00  & 8 &  1.43  &  0.76  &  40.10  \\
EGARCH--$ \mathcal{T}$ & 9 &  0.26  &  1.00  & 10 &  1.62  &  0.55  &  40.30  \\
EGARCH--$ \mathcal{N}$ & 10 &  0.32  &  1.00  & 9 &  1.62  &  0.55  &  40.40  \\
TGARCH--$ \mathcal{N}$ & 11 &  0.39  &  1.00  & 11 &  1.64  &  0.52  &  40.40  \\
GARCH--$ \mathcal{N}$ & 12 &  0.42  &  1.00  & 12 &  1.65  &  0.49  &  40.40  \\
CGARCH--$ \mathcal{N}$ & 13 &  0.59  &  1.00  & 13 &  1.73  &  0.41  &  40.50  \\
TGARCH--$ \mathcal{T}$ & 14 &  0.67  &  0.99  & 14 &  1.79  &  0.30  &  40.60  \\
NGARCH--$ \mathcal{N}$ & 15 &  0.96  &  0.93  & 15 &  1.90  &  0.19  &  40.80  \\
CGARCH--$ \mathcal{T}$ & 16 &  1.08  &  0.87  & 16 &  2.00  &  0.11  &  40.80  \\
GARCH--$ \mathcal{T}$ & 17 &  1.11  &  0.85  & 17 &  2.00  &  0.10  &  40.90  \\
NGARCH--$ \mathcal{T}$ & 18 &  1.40  &  0.61  & 18 &  2.14  &  0.01  &  41.10  \\
\hline
\multicolumn{8}{l}{\textit{SXW1E: 7 eliminations}}\\
GJRGARCH--$ \mathcal{T}$ & 1 & -1.61  &  1.00  & 1 & -0.36  &  1.00  &  28.80  \\
APARCH--$ \mathcal{T}$ & 2 & -1.21  &  1.00  & 2 &  0.36  &  1.00  &  29.00  \\
EGARCH--$ \mathcal{T}$ & 3 & -0.97  &  1.00  & 3 &  0.54  &  1.00  &  29.10  \\
TGARCH--$ \mathcal{T}$ & 4 & -0.79  &  1.00  & 4 &  0.67  &  1.00  &  29.20  \\
APARCH--$ \mathcal{N}$ & 5 & -0.37  &  1.00  & 5 &  0.93  &  0.98  &  29.30  \\
GJRGARCH--$ \mathcal{N}$ & 6 & -0.33  &  1.00  & 6 &  0.96  &  0.97  &  29.30  \\
TGARCH--$ \mathcal{N}$ & 7 &  0.06  &  1.00  & 7 &  1.16  &  0.90  &  29.40  \\
AVGARCH--$ \mathcal{T}$ & 8 &  0.28  &  1.00  & 8 &  1.31  &  0.76  &  29.50  \\
EGARCH--$ \mathcal{N}$ & 9 &  0.78  &  0.94  & 9 &  1.56  &  0.42  &  29.70  \\
NGARCH--$ \mathcal{T}$ & 10 &  0.85  &  0.91  & 10 &  1.62  &  0.33  &  29.70  \\
GARCH--$ \mathcal{T}$ & 11 &  0.90  &  0.88  & 11 &  1.65  &  0.29  &  29.80  \\
AVGARCH--$ \mathcal{N}$ & 12 &  1.13  &  0.64  & 12 &  1.76  &  0.15  &  29.90  \\
CGARCH--$ \mathcal{T}$ & 13 &  1.21  &  0.53  & 13 &  1.82  &  0.09  &  29.90  \\
\hline
\multicolumn{8}{l}{\textit{SXXP: 15 eliminations}}\\
TGARCH--$ \mathcal{T}$ & 1 &  1.25  &  1.00  & 1 &  0.35  &  1.00  &  36.60  \\
AVGARCH--$ \mathcal{T}$ & 2 &  0.67  &  1.00  & 2 &  0.35  &  1.00  &  36.70  \\
EGARCH--$ \mathcal{T}$ & 3 &  0.12  &  1.00  & 3 &  0.75  &  0.99  &  36.80  \\
APARCH--$ \mathcal{T}$ & 4 &  0.85  &  0.78  & 4 &  1.33  &  0.32  &  37.00  \\
AVGARCH--$ \mathcal{N}$ & 5 &  1.16  &  0.19  & 5 &  1.46  &  0.17  & 37.10 \\
\bottomrule
\end{tabular}}
\caption{\footnotesize{Comparison of the SSMs for the four considered international stock indexes. The p--values of the $\mathrm{T}_{\rm R,M}$ and $\mathrm{T}_{\max ,\rm M}$ statistics, are reported in the third and seventh columns, respectively. The p--value of the test statistic, is equal to the minimum of the overall p--values. The columns $\mathrm{Rank}_{\mathrm{R,M}}$ and $\mathrm{Rank}_{\mathrm{max,M}}$ report the ranking over the models belonging to the SSMs. Finally, the last column Loss$\times10^3$ is the average loss across the considered period.}}
\label{tab:MCS_composition}
\end{table}

\begin{table}[!ht]
\centering
\begin{tabular}{lccccccc}
\toprule
  & \multicolumn{3}{c}{$\mathrm{VaR_{Dyn}}$} & \multicolumn{3}{c}{$\mathrm{VaR_{Avg}}$}\\
  \cmidrule(lr){1-1} \cmidrule(lr){2-4}\cmidrule(lr){5-7}
 Asset & AE & ADmean & ADmax & AE & ADmean & ADmax \\
 \cmidrule(lr){1-1}\cmidrule(lr){2-4}\cmidrule(lr){5-7}
SXA1E &  1.55  &  0.698  &  3.121  &  2.05  &  0.700  &  3.315 \\
SXP1E &  1.10  &  0.950  &  3.787  &  1.45  &  0.798  &  3.766 \\
SXW1E &  1.65  &  0.420  &  1.974  &  2.30  &  0.426  &  2.051 \\
SXXP &  1.95  &  0.514  &  2.520  &  2.50  &  0.515  &  2.836 \\
\bottomrule
\end{tabular}
\caption{\footnotesize{VaR backtesting measures of the dynamic VaR combination $\rm VaR_{Dyn}$ and the static average $\rm VaR_{Avg}$.}}
\label{tab:var_comparison}
\end{table}
\indent For each index, Table \ref{tab:MCS_composition} reports the compositions of the SSM. Naturally, the higher the number of eliminated models, the higher the heterogeneity of the competing forecasts. On the contrary, if the final SSM contains a big portion of the starting ${\rm M}^0$ set, then the competing model are statistically equivalent in term of their forecast ability of future VaR levels. The p--values of the $\mathrm{T}_{\rm R,M}$ and $\mathrm{T}_{\max ,\rm M}$ statistics, are reported in the second and sixth columns, respectively. The p--value of the test statistic, is equal to the minimum of the overall p--values reported in the fourth and seventh columns of Table \ref{tab:MCS_composition}, respectively. For a detailed discussion about the interpretation of the MCS p--values we refer to \cite{hansen_etal.2011}.
The estimated SSMs differ for the number of the eliminated models as well as for their compositions. We can observe that, for the SXP1E index, only 2 models were eliminated by the MCS procedure. This empirical finding highlights the statistical equivalence of forecasting future VaR levels using a simple model such as the GARCH(1,1)--$\mathcal{N}$ or a more sophisticated one like the GAS--$\mathcal{T}$. Furthermore, this evidence suggests that, the SXP1E index may not be affected by some stylised facts such as the leverage effect or by complex nonlinear conditional volatility dynamics. For the SXP1E and SXW1E indexes the MCS procedure eliminates five and seven models, respectively. An higher level of discrimination among models is instead evident for the SXXP index. In fact, in that case, 15 of the 20 considered models do not belong to the final SSM and the five remaining models are characterised by strong nonlinear dynamics for the conditional volatility process. Moreover, 4 of the 5 remaining models are characterised by a Student--t distribution, meaning that, for the SXXP index, a more leptokurtic conditional distribution is required to improve future VaR forecasts. At a first glance, it would seem strange to observe that the composition of the SSM is quite homogeneous with respect to the conditional distribution assumption. Indeed, apart from the SXXP index, all series appear to be well described by a Gaussian, as well as by a Student--t distribution. However, as discussed by \cite{jondeau_etal.2007}, it should be noted that the Gaussian assumption for the innovations does not implies gaussianity for the unconditional distribution of the returns. Concerning the empirical relevance of the distribution assumption, the considered return series can be viewed as highly diversified portfolios. Well diversified portfolios are characterised by the fact that positive and negative tail events affecting the conditional distribution and its kurtosis, are mitigated by the diversification. The need for an heavy--tailed distribution to describe the SXXP return index is probably due to the higher impact that the GFC had to the European economy as compared to the other countries. Finally, it is interesting to note that, all the CAViaR specifications are always excluded from the SSM, suggesting the CAViaR dynamics do not adequately describe future VaR levels.\newline
\indent In order to test the benefits of using the MCS procedure, we also compare two different VaR aggregation techniques. The first aggregation technique, $\mathrm{VaR_{avg}}$, simply averages the VaR forecasts delivered by the 22 competing models, while the second one, $\mathrm{VaR_{Dyn}}$, is instead performed by applying the dynamic VaR combination method proposed in \cite{bernardi_etal.2014} to the VaR forecasts delivered by the MCS models. The dynamic VaR combination technique, instead, averages VaR forecasts using a dynamic updating rule on each model's relative contribution to the total quantile loss. Table \ref{tab:var_comparison} reports the backtesting performances of the two VaR aggregation methods. Three VaR backtesting measures are considered. The first is the Actual over Expected ratio AE, defined as the ratio between the realised VaR exceedances over a given time horizon and their \qmo a priori\qmcsp expected values; VaR forecasts series for which the AE ratio is closer to the unity are preferred. The second and the third backtesting measures are the mean and maximum Absolute Deviation (ADmean and ADmax) of VaR violating returns described in \cite{mcaleer_daveiga.2008}. The ${\rm AD}$ in general provides a measure of the expected loss given a VaR violation; of course models with lower mean and/or maximum ADs are preferred. As showed in Table \ref{tab:var_comparison}, except for the SXP1E index, the $\mathrm{VaR_{Dyn}}$ series always report lower ADmean and ADmax compared with the $\mathrm{VaR_{avg}}$, while the AE ration is strongly improved for all the considered indexes.
\section{Conclusion}
\label{sec:conclusion}
%
\noindent This paper compares alternative model specifications in term of their VaR forecasting performances. The model comparison is performed using the MCS procedure recently proposed by \cite{hansen_etal.2011}. The MCS technique is particularly useful when several different models are available and it is not obvious which one performs better. The MCS sequence of tests delivers the Superior Set of Models having Equal Predictive Ability in terms of an user supplied loss function discriminating models. This flexibility helps to discriminate models with respect to desired characteristics, such as, for example, their forecasting performances. In our empirical application, we compare the VaR forecast ability of several models.
More specifically, the direct quantile estimates obtained by the dynamic CAViaR models of \cite{engle_manganelli.2004} are compared with the VaR forecast delivered by several ARCH--type models of \cite{engle.1982} and with those obtained by two different specifications of the Generalised Autoregressive Score (GAS) models of \cite{creal_etal.2013} and \cite{harvey.2013}. The MCS procedure is firstly performed to reduce the initial number of models, and then to show that accounting for the VaR dynamic model averaging technique of \cite{bernardi_etal.2014} improves the VaR forecast performance.
Our empirical results, suggest that, after the Global Financial Crisis (GFC) of 2007--2008, highly non--linear volatility models are preferred by the MCS procedure for the European countries. On the contrary, quite homogenous results, with respect to the models' complexity, were found for the the North America and Asia Pacific regions.
\indent The paper also illustrates the main features of the provided \textsf{R} package \texttt{MCS} available on the \texttt{CRAN} repository, \url{http://cran.r-project.org/web/packages/MCS/index.html}. The \texttt{MCS} package is very flexible since it allows for the specification of the model's types and loss functions that can be supplied by the user. This freedom allows for the user to concentrate on substantive issues, such as the construction of the initial set of model's specifications $\mathrm{M}^{0}$, without being limited by the the software constrains.\newline
\section*{Acknowledgments}
%
\noindent This research is supported by the Italian Ministry of Research PRIN 2013--2015, ``Multivariate Statistical Methods for Risk Assessment'' (MISURA), and by the ``Carlo Giannini Research Fellowship'', the ``Centro Interuniversitario di Econometria'' (CIdE) and ``UniCredit Foundation''. In the development of package \texttt{MCS} we have benefited from the suggestions and help of several users. In particular, we would like to thank Riccardo Sucapane for his constructive comments on previous drafts of this work. Our sincere thanks go to all the developers of \textsf{R} since without their continued effort and support no contributed package would exist.
%

%
\bibliographystyle{apalike}
\bibliography{references}
\end{document}